# Sequential Phase Transitions and Transient Structured Fluctuations in Two-Dimensional Systems with a High-Density Kagome Lattice Phase

Linsey Nowack and Stuart A. Rice

*James Franck Institute, Department of Chemistry, and the Chicago Center for Theoretical Chemistry, The University of Chicago, Chicago, IL, 60637*

## Abstract

We report the results of molecular dynamics simulation studies that explore two features of the phase diagrams of two two-dimensional systems composed of particles with everywhere repulsive isotropic pair potentials, one proposed by Piñeros, Baldea, and Truskett, and the other by Zhang, Stillinger, and Torquato, each of which supports a high-density Kagome lattice phase. These features are (i) the sequences of phases and the phase transitions characteristic of each system as the density is increased along an isotherm, and (ii) the character of transient structured fluctuations in the phases adjacent to a high-density Kagome lattice phase. As to (i), comparison of the sequences of phases supported by the pair potentials used in our simulations and those supported by other pair potentials provides information vis a vis a relationship between the shape of the pair potential and the density dependence of the sequence of phases. The commonalities in the phase diagrams of the several 2D systems suggests the existence of a universal mechanism driving all to favor a similar series of packing arrangements as the density is increased. However, the collection of simulations considered shows that satisfying the only such general rule proposed, namely the Süto theorem relating the character of the Fourier transform of the pair potential to the existence of multiple ground states of a system, is not a necessary condition for the support of multiple distinct lattice structures by a particular pair potential. As to (ii), the "open" structure of a Kagome lattice requires an unusual rearrangement of the particle packing in the phase from which it emerges. We find that on an isotherm in the liquid phase, close to the liquid-to-Kagome phase transition, the transient structured fluctuations in the liquid have Kagome symmetry whereas deeper in the liquid phase the transient structured fluctuations have hexagonal symmetry. As the deviation of the liquid density from the transition density decreases transient fluctuations with hexagonal symmetry are replaced with those with Kagome symmetry with a coexistence domain of a few percent. When the transition is string phase-to-Kagome phase the transient structured density fluctuations in the string phase near the transition do not have Kagome character; there are both configurations with six-fold and other than six-fold symmetries, with stronger preference for six-fold symmetry in the Truskett system than in the Torquato system. The path of the string-to-Kagome transition in the Truskett system involves intermediate particle configurations with honeycomb symmetry that subsequently buckle to form a Kagome lattice. The path of the string-to-Kagome transition in the Torquato system suggests that the Kagome phase is formed by coiled strings merging together at three-particle joining sites. The increasing concentration of these joining sites as the density is increased generates a Kagome phase with imperfections such as 8-particle rings.

## Introduction

Over the last few decades, computer simulations have revealed that a two-dimensional (2D) system composed of particles whose direct interaction is a monotonic isotropic everywhere repulsive pair potential can support a variety of intricate ordered phases. The structures of these phases include lattice symmetries that are triangular (hexagonal), square, honeycomb, Kagome, sigma, chain (a.k.a. "string" or "stripe"), and more [1-7]. The support of so many different ordered structures, stable in different density regimes, calls attention to the complexity of the entropic contribution to the effective interaction in these systems. We expect that entropic contribution, which has the character of a many-body interaction, must be very sensitive to the system density, and to the shape of the direct pair potential. This paper addresses four questions related to these expectations:

1. As a function of increasing density, do different isotropic monotonic repulsive pair potentials support the same phases, phase transitions and sequence of phases?
2. What are the qualitative relationships between the shape of the direct pair potential and the sequence of phases supported?
3. Does the spectrum of transient ordered fluctuations in the liquid phase of a system composed of particles whose direct interaction is a monotonic soft isotropic repulsive pair potential that supports many different ordered solid phases, differ from that in the hard disk liquid?

4. What transient ordered fluctuations in the liquid are associated with a transition from the liquid state to a solid with an open lattice, e.g. a Kagome lattice?

Question (1) is stimulated by the results of investigations of the phase diagrams of 2D systems with different everywhere repulsive central pair interactions. The results reported in references [3-7] and those obtained in this investigation reveal both similarities and differences in the sequence of phases and phase transitions supported as a function of isothermally increasing density. Given the centrosymmetric character of all of the sampled pair interactions, the appearance in the phase diagram of density regimes with hexagonal and square packed phases, as well as the more open honeycomb and Kagome packed phases, is arguably not surprising. What is, in our opinion, more unusual is the appearance in the phase diagram of density regimes with phases composed of pairs of particles, and of strings of particles. It is noteworthy that, as a function of the pair potential, the sequence of phases encountered as the density is increased along an isotherm exhibits both common and distinctive behavior. For the sampling of pair potentials we have considered the lowest density phase transition is always from the liquid to a hexagonal phase and the highest density ordered phase found is also hexagonal. However, some of the pair potentials support a set of sequential transitions that involves the structures pairs→ strings → Kagome → hexagonal, while others support a more complex sequence that can include pentagonal and stretched hexagonal packings. The role played by the shape and range of the monotonic repulsive potential in determining the observed sequences of phase transitions (question (2)) is incompletely understood.

Answers to questions (3) and (4) are related via our definition of a transient ordered fluctuation. Briefly put, we identify an ordered transient fluctuation as a symmetric configuration of a small number of particles whose diffraction pattern consists of discrete Bragg peaks. Each transient ordered fluctuation has a unique signature that distinguishes it from other transient structured fluctuations in the liquid phase. These signatures are determined from the aperture cross correlation function (ACCF) developed by Ackerson and Clark [8]. The ACCF correlates the appearance, at the same time or as a function of time difference, of two Bragg peaks in the radiation scattered from fluid contained within a small aperture. The scattered radiation is monitored by two detectors that can be chosen to probe the same or different values of the magnitude and direction of the scattering vector. We are concerned in this paper with the same time cross correlation function. The utility of the ACCF is that it *discovers* an explicit signature of local order in the liquid phase instead of assigning order based on an arbitrarily assigned measure of the particle configurations' nearness to a preselected structure [9].

It is reasonable to expect that a 2D liquid of particles with a direct pair interaction that is monotonic repulsive is capable of exhibiting many different types of transient structured fluctuations. This expectation follows from two observations. First, the average number density around a particle in the liquid, described by the radial distribution function, shows that the radial density distribution is inhomogeneous. Instantaneous realizations of the distributions of particles around a central particle deviate from the mean described by the radial distribution function; the deviations in number density are comparable in magnitude with the magnitude of the inhomogeneities in the average radial density [10], implying that there is ample accessible configuration space for many different transient orderings of the particles. Second, the mostly entropy driven self-assembly of repulsive particles into an ordered configuration requires a delicate balance in the multiple interparticle interactions and spatial arrangements. Given that fluctuations in the liquid are large and the relationship between the favored structure and pair potential delicate, it is reasonable to expect to find numerous transient structured fluctuation motifs within the liquid, including some not associated with the ordered solid structures of the system.

We have previously reported studies of the occurrence of transient ordered fluctuations in the liquid phase of the hard disc system [11] and the liquid phase of a 2D system with a pair potential designed by Dudalov et al [13-15], the latter known to support two hexagonal solid phases, a square solid phase, and a 12-fold ordered phase [15]. In both systems our calculations revealed evidence of neighboring particles fleetingly favoring ordered configurations both adjacent to phase transition lines and when the system is deep within the liquid phase. Moreover, these structured fluctuations exhibited the same symmetries as the ordered phases in the system; no other symmetries were observed to occur. Because the phase diagrams of the systems studied have only close-packed solid phases, the character of transient ordered fluctuations close to a transition into an ordered phase with an "open" lattice has not been studied. In this paper we explore the existence of transient structured fluctuations in a 2D liquid phase that shares a transition boundary with a Kagome phase, and in a 2D string phase that shares a transition boundary with a Kagome phase, to learn how ordered fluctuations in a nominally close packed phase develop the open ordered configuration characteristic of the destination solid.

## Systems and Methods

1. **Truskett and Torquato Model 2D Systems**

In this study, molecular dynamic simulations of 2D systems with periodic boundary conditions containing, respectively, 1200 particles and 11250 particles, were performed using Sandia National Laboratory's LAMMPS package [16]. Systems with two different pair interactions were simulated: one consisting of particles interacting with a pair potential designed by Truskett et al [1] and another consisting of particles interacting with a pair potential designed by Torquato et al [2]. For the sake of clarity, these pair potentials will be subsequently referred to as the Truskett potential and the Torquato potential.

The Truskett potential has the form [1]

$$u_1\left(\frac{r}{\sigma}\right) = \epsilon\left\{A\left(\frac{r}{\sigma}\right)^{-n} + \sum_{i=1}^{2}\lambda_i\left(1 - \tanh\left[k_i\left(\frac{r}{\sigma} - \delta_i\right)\right]\right)\right\} H[(r_{cut} - r)/\sigma] \qquad (1)$$

where $\epsilon$ and $\sigma$ represent the energy and length scales, H is the Heaviside function, $r_{cut}$ is the cutoff radius (=3.00000), and the rest of the parameters have the values A= 0.01978, n = 5.49978, $\lambda_1$ = -0.06066, $k_1$ = 2.53278, $\delta_1$ = 1.94071, $\lambda_2$ = 1.06271, $k_2$ = 1.73321, $\delta_2$ = 1.04372). The Torquato potential has the form [2]

$$u_2\left(\frac{r}{\sigma}\right) = \epsilon\left\{\left(b\left(\frac{r}{\sigma}\right)^{-12} + c_0 + c_1\left(\frac{r}{\sigma}\right)\right)\left(\frac{(r-r_{cut})}{\sigma}\right)^2\right\} H[(r_{cut} - r)/\sigma] \qquad (2)$$

with b = 5.9860 × 10$^{-2}$, $c_0$ = −1.2811, $c_1$ = 2.1521, and $r_{cut}$ = 2.0364 [2]. Throughout the remainder of this paper we use the reduced units r* ≡ r/σ, P* ≡ Pσ²/ε, V* ≡ V/Nσ², $\rho^* = \frac{1}{V^*}$, τ* ≡ (mσ²/ε)$^{1/2}$, and T* ≡ $k_B$T/ε. And, as we deal only with the representation in reduced variables, from here onwards the asterisks will be omitted. The Truskett and Torquato pair potentials and their corresponding pair forces are plotted together in Figure 1. Both potentials are monotonic repulsive, and divergent at the origin. Each has a small shoulder; the shoulder of the Torquato potential is more pronounced than that of the Truskett potential. The pair force generated by the Torquato potential has a shallow minimum at about r = 1 whereas that generated by the Truskett potential has a gently sloping plateau at about r = 1.

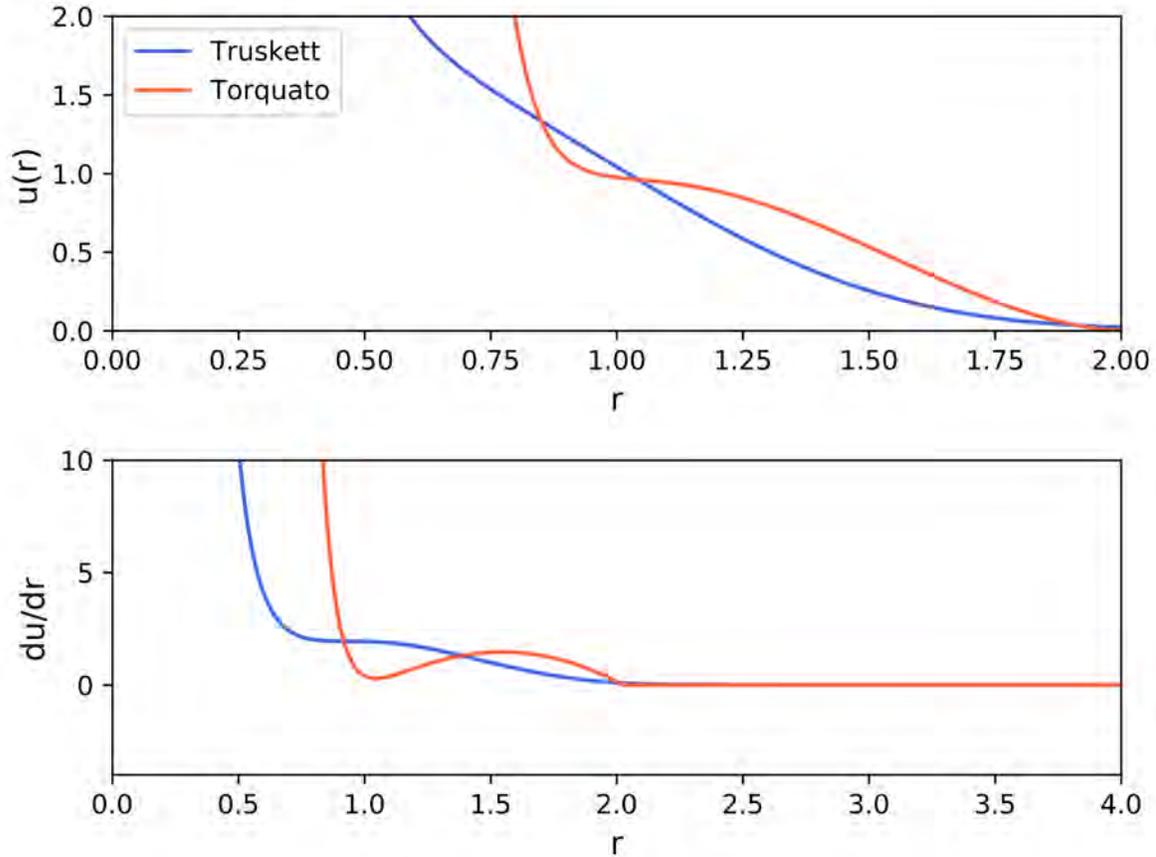

**Figure 1.** The Truskett and Torquato potential functions (top) and force functions (bottom).

We have carried out simulations at constant temperature and density along the isotherms T = 0.005, T = 0.010, T = 0.020, T = 0.023, and T = 0.030. For the Truskett system, the densities of the simulations ranged from ρ = 0.100 to ρ = 1.600, and for the Torquato system the densities of the simulations ranged from ρ = 0.100 to ρ = 1.100. These density ranges were chosen such that in both systems the phase diagram included the high-density transition out of the Kagome phase. Our execution of the simulations included a six-step process of cooling according to the following scheme:

1) Equilibrate at high temperature for 500,000 timesteps;
2) Cool from the high temperature to an intermediate temperature over 500,000 timesteps;

3) Equilibrate at the intermediate temperature for 500,000 timesteps;
4) Cool from the intermediate temperature to a target temperature over 500,000 timesteps;
5) Equilibrate at the target temperature for $10^6$ timesteps;
6) Data collection over $10^6$ timesteps.

The timestep used was 0.001 τ. For the T = 0.005 isotherm simulations, the high temperature was T = 0.023, the intermediate temperature was T = 0.010, and the target temperature was T = 0.005. The corresponding temperatures for the other simulations are displayed in Table 1.

| Isotherm | T = 0.005 | T = 0.010 | T = 0.020 | T = 0.023 | T = 0.030 |
| --- | --- | --- | --- | --- | --- |
| High temperature | T = 0.023 | T = 0.023 | T = 0.023 | T = 0.023 | T = 0.030 |
| Intermediate temperature | T = 0.010 | T = 0.010 | T = 0.020 | T = 0.023 | T = 0.030 |
| Target temperature | T = 0.005 | T = 0.010 | T = 0.020 | T = 0.023 | T = 0.030 |

**Table 1.** The high, intermediate and target temperatures used in the cooling process for each simulation. Note the dependence on the choice of target temperature.

The cooling scheme described was chosen to imitate that used in Truskett et al's 2016 paper. In that paper a Monte Carlo method was employed to simulate a two-dimensional Kagome lattice phase at T = 0.005 and ρ = 1.400. Truskett and coworkers' system consisted of 1200 particles initially equilibrated at T = 0.023 before being quenched at constant density to T = 0.010 and then to T = 0.005 [1]. This procedure is mimicked in our T = 0.005 isotherm simulations. For higher temperature isotherms, we used a different approach. As indicated by the data in Table 1, for the T = 0.030 isotherm, for example, the simulations were equilibrated at T = 0.030 for 3,000,000 timesteps before data were collected over another 1,000,000 timesteps. The same process was used for the T = 0.023 isotherm simulations. In all of the simulation runs, the total energy was monitored to ensure that the system reached equilibrium before data collection was started. The coordinates of the particles were recorded every 100 timesteps and displayed using the program Visual Molecular Dynamics (VMD) [17].

## 2. Methods of Analysis

The analysis of the data accumulated in the several simulations included the shape of the pressure-density isotherms, calculation of the diffraction patterns and structure functions for the equilibrium particle configurations, calculation of ACCFs for selected temperature-density points, and visual examination of particle configurations. Instantaneous pressure-density isotherms were calculated every 1000 timesteps using the virial theorem expression and taking account of the number of images of atoms outside yet close to the periodic boundaries of the simulation cell. The instantaneous pressures for the full $10^6$-timestep data collection were averaged together to produce the pressure-density isotherms. An example, the T = 0.005 isotherm plot for the 1200 particle system with Truskett pair interaction (Eq. (1)), is displayed in Fig. 2.

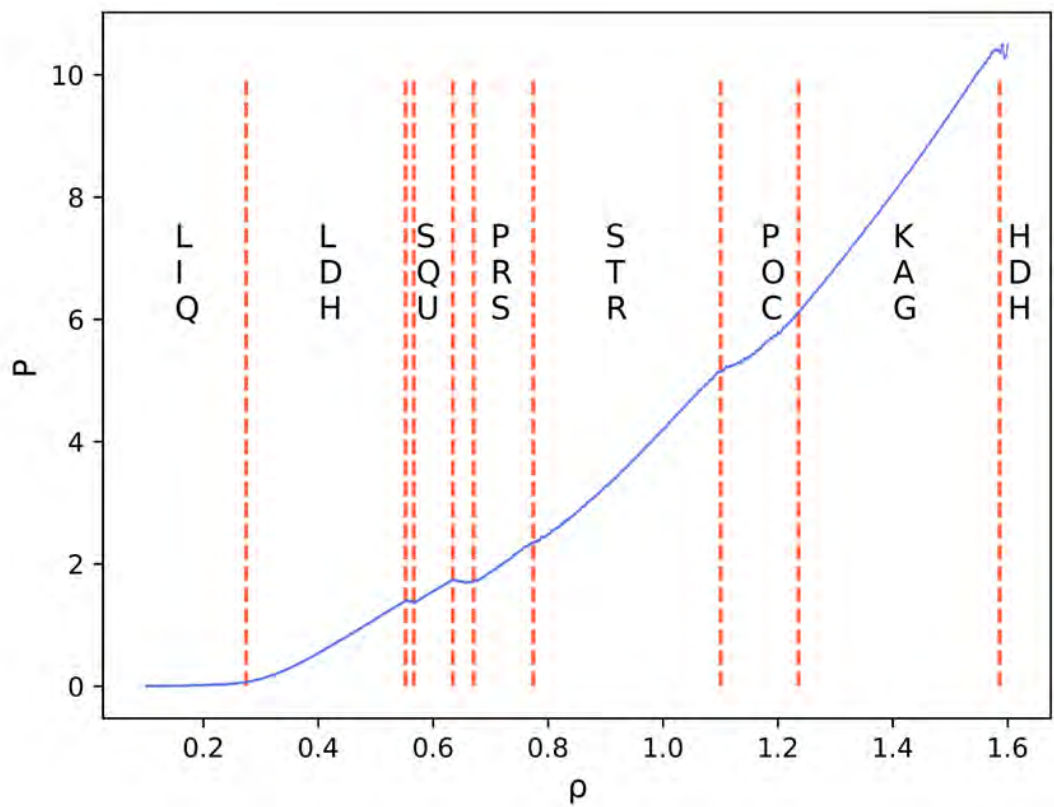

**Figure 2.** Equation of state of a 2D 1200 particle system with Truskett pair interaction along a low-temperature (T = 0.005) isotherm. LIQ = liquid, LDH = low-density hexagonal solid, SQU = square solid, PRS = pairs, STR = string, POC = pockets, KAG = Kagome, HDH = high-density hexagonal solid.

The pressure-density plot displayed in Fig. 2 clearly suggests that that under isothermal compression the system passes through a sequence of phases; the locations of phase transitions signaled by the changes in shape of the isotherm are indicated by the vertical lines. Our interpretation of the simulation data relies on use of the calculated structure function, the image of the corresponding diffraction pattern and, for the liquid phase, the ACCF of the system.

The structure function, S(**q**), describes the intensity of radiation scattered from the system that is monitored with a single detector. We have computed S(**q**) for our 2D systems from the time-averaged instantaneous scattered intensity with wave vector **q**, I(**q**,t),

$$S(\mathbf{q}) = \frac{1}{N}\langle I(\mathbf{q}, t)\rangle, \tag{4}$$

$$I(\mathbf{q}, t) = \sum_{ij}^{N} \cos[\mathbf{q} \cdot \{\mathbf{r_i}(t) - \mathbf{r_j}(t)\}] \quad, \tag{5}$$

using the last 100,000 timesteps of each simulation. Particle coordinates were sampled every 100 timesteps and converted into an image in pixels. The squared absolute value of the Fourier transform of each image was then taken to produce the corresponding instantaneous structure function, with subsequent averaging of the instantaneous structure functions as in Eq. (4). In Eq. (5) the sum is over all of the particles in the simulation cell.

The same-time normalized ACCF of a system, defined by

$$C(\mathbf{k}, \mathbf{q}) = \frac{\langle I(\mathbf{k})I(\mathbf{q})\rangle}{\langle I(\mathbf{k})\rangle\langle I(\mathbf{q})\rangle}, \tag{6}$$

provides information not contained in S(**q**). In Eq. (6), the average is over particles contained within a small aperture and the scattered intensity is measured with two detectors located to monitor scattered radiation with wave vectors **q** and **k**. When one detector is kept stationary and |**k**| = |**q**|, the cross correlated intensity with respect to angle takes the form

$$C(\varphi) = \frac{\langle I(\varphi_0)I(\varphi_0+\varphi)\rangle}{\langle I(\varphi_0)\rangle\langle I(\varphi_0+\varphi)\rangle} \qquad (7)$$

with $\varphi_0$ constant corresponding to the fixed direction of the detector at **k** and $(\varphi_0 + \varphi)$ the angle between the detectors at **k** and **q** [8]. Choosing $|\mathbf{k}| = |\mathbf{q}|$ implies that the structure of an ordered fluctuation can be inferred from the intensity distribution in one ring of the diffraction pattern; we have previously chosen $|\mathbf{q}|$ to correspond to the first diffraction ring [12,13]. That choice works well for simple ordered structures, such as hexagonal and square, but not for distributions with symmetries related to more complex lattices. For example, the diffraction patterns of both the hexagonal and Kagome lattices (see Fig. 3) have first diffraction rings with six spots of equal intensity, hence an ACCF that is limited to the representation of the first ring intensity distribution cannot distinguish between transient hexagonal and transient Kagome ordering. The intensities of the spots in the second rings of the diffraction patterns of these lattices differ; they are all the same for the hexagonal lattice whereas for the Kagome lattice the intensities at the second ring's six vertices are larger than the intensities halfway along the lines between the vertices. To uniquely identify an ordered fluctuation with Kagome structure we compute the ACCF for the values of **q** and **k** that characterize the vertices and the halfway points along the lines connecting the vertices in the second ring of the diffraction pattern. Comparison of these ACCFs, computed simultaneously for the same aperture, reveals the modulation and angular distributions of diffraction intensity that characterize the structure of the ordered fluctuation. For all of the ACCF calculations described in this paper the apertures chosen were illuminated at individual timesteps such that, over the timespan of the equilibrated simulation, the total area of the simulation box is illuminated at least once. The aperture from which the scattered intensity was collected was chosen to be circular with a radius $4\sigma$, which is comparable to or larger than the correlation length for the structured fluctuations.

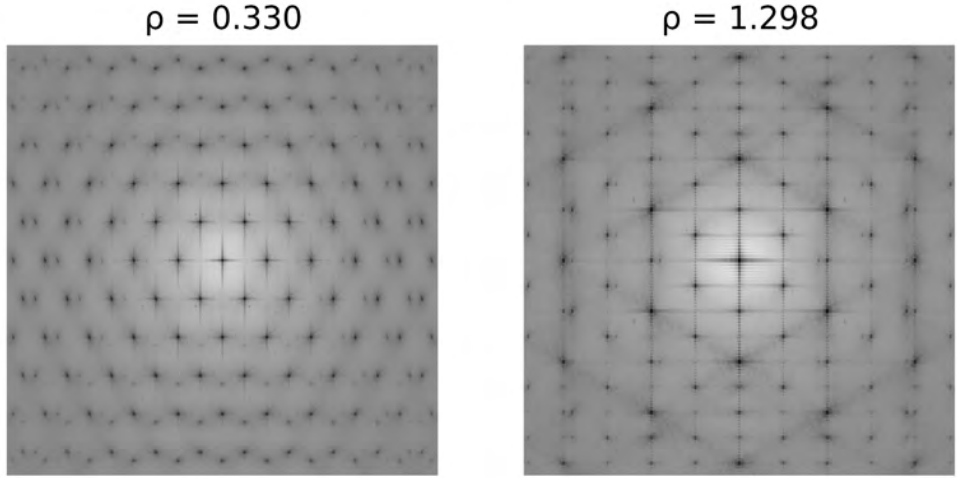

**Figure 3.** The low-density hexagonal solid structure function (left) and the Kagome lattice structure function (right) for the T = 0.005 Truskett system of 1200 particles. Note the different distributions of diffraction spot intensities in the second diffraction rings of these two patterns.

## Results

### 1. 2D Phases Supported by the Truskett Pair Interaction

Figures 4 and 5 display snapshots of the Truskett system particle configurations for several densities along the low-temperature isotherm $T = 0.005$. These images, when correlated with changes in slope and loops in the pressure-density isotherm, clearly suggest that isothermal compression of a 2D system with Truskett interaction generates a sequence of phases with phase boundaries shown by the vertical lines in Fig. 2. Along the isotherm $T = 0.005$ the sequence of phases we have identified is:

Liquid → Low-Density Hexagonal → Square → Pairs → String → Kagome → High-Density Hexagonal.

Acknowledging that the pressure-density isotherm displayed in Fig. 2 has both subtle and obvious slope changes and loops, we associate particular phases with the density domains delineated by the vertical lines. The lowest transition apparent from the shape of the pressure-density isotherm is a smooth transformation from the liquid to a low-density hexagonal solid, at about $\rho = 0.3$. The low-density hexagonal solid phase is stable until about $\rho = 0.552$ at which density coexistence with a square packed solid phase commences, the signature of which is the loop in the pressure-density isotherm between $\rho = 0.552$ and $\rho = 0.566$; the phase coexistence in this density range suggests that the low-density hexagonal to square solid transformation is a first-order transition. Similarly, between $\rho = 0.634$ and $\rho = 0.668$ there is a first order transition from the square solid phase to a phase of coupled particles, or "pairs" (note the coexistence domain). The centers of mass of the pairs form a stretched hexagonal lattice (Figs. 6 and 7).

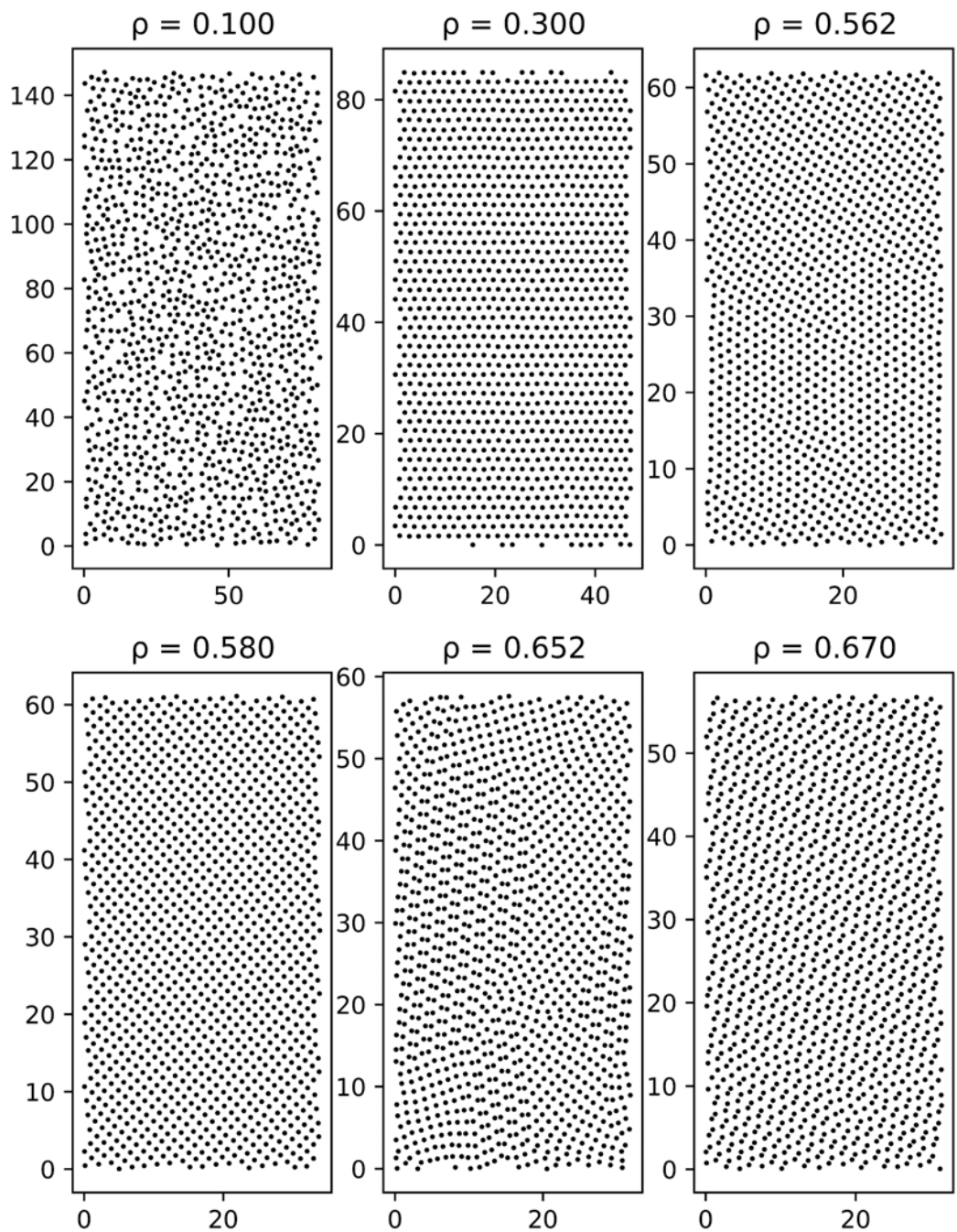

**Figure 4.** Snapshots of the particle configurations in the Truskett system with 1200 particles at different densities with T = 0.005.

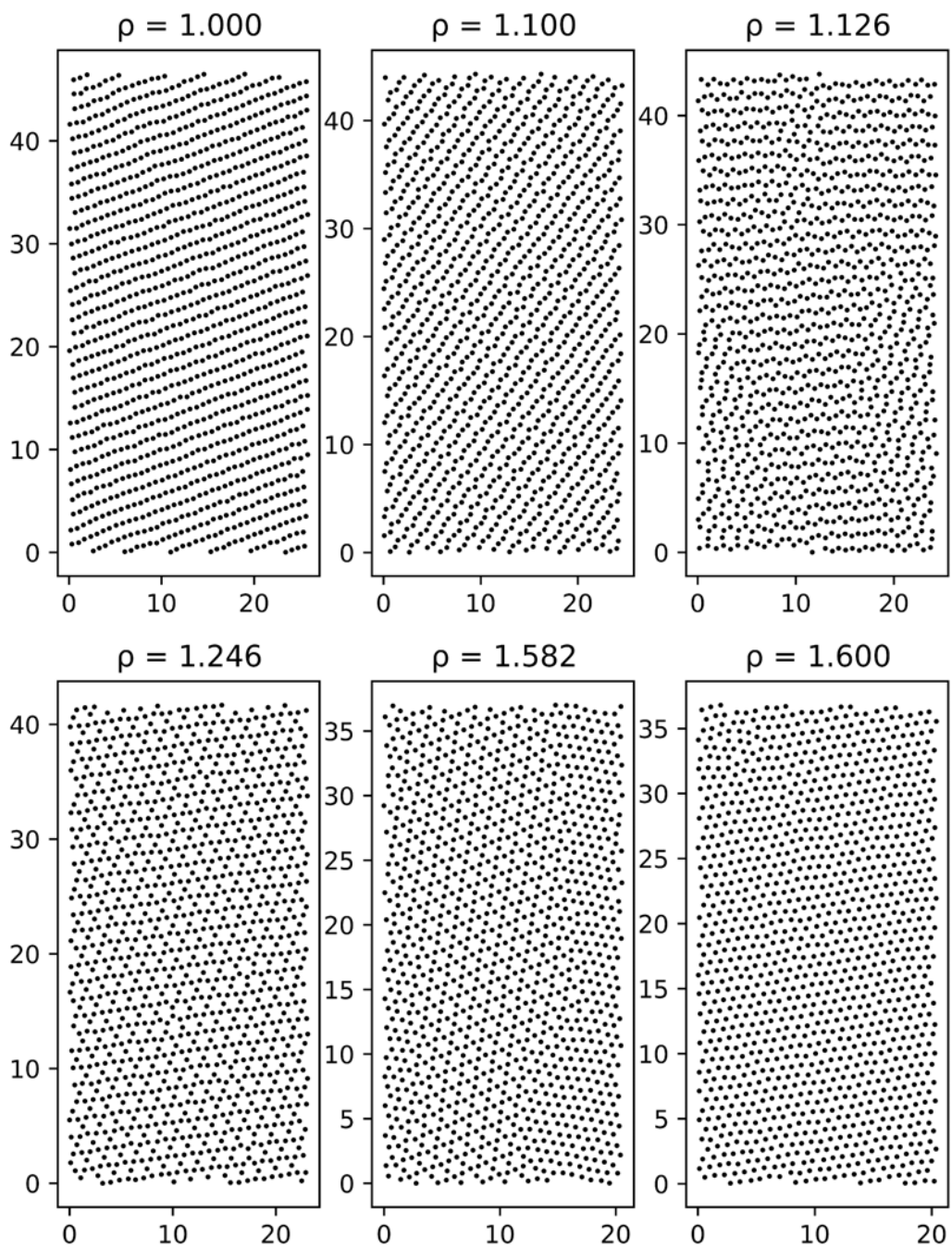

**Figure 5.** Snapshots of the particle configurations in the Truskett system with 1200 particles at different densities with T = 0.005.

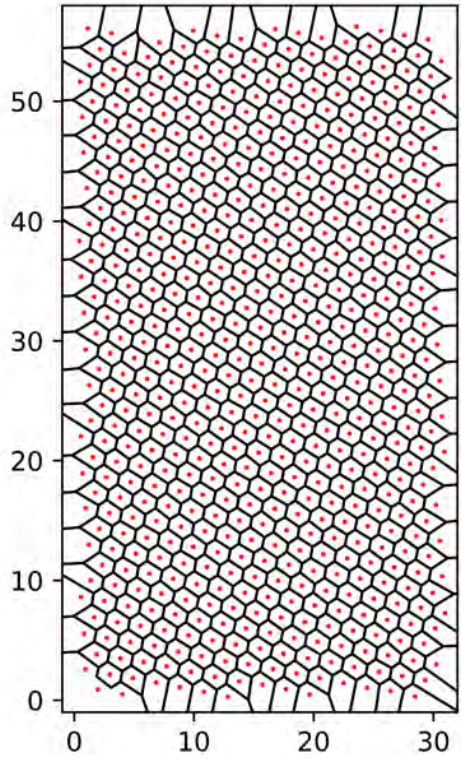

**Figure 6.** Voronoi tessellation of the particle configuration in the pairs phase, ρ = 0.678, T = 0.005, of the Truskett system with 1200 particles. The red dots mark the positions of the centers of mass of the pairs such as seen the lower right panel of Fig. 4. Note the stretching of the hexagonal cells along the NW-SE axis.

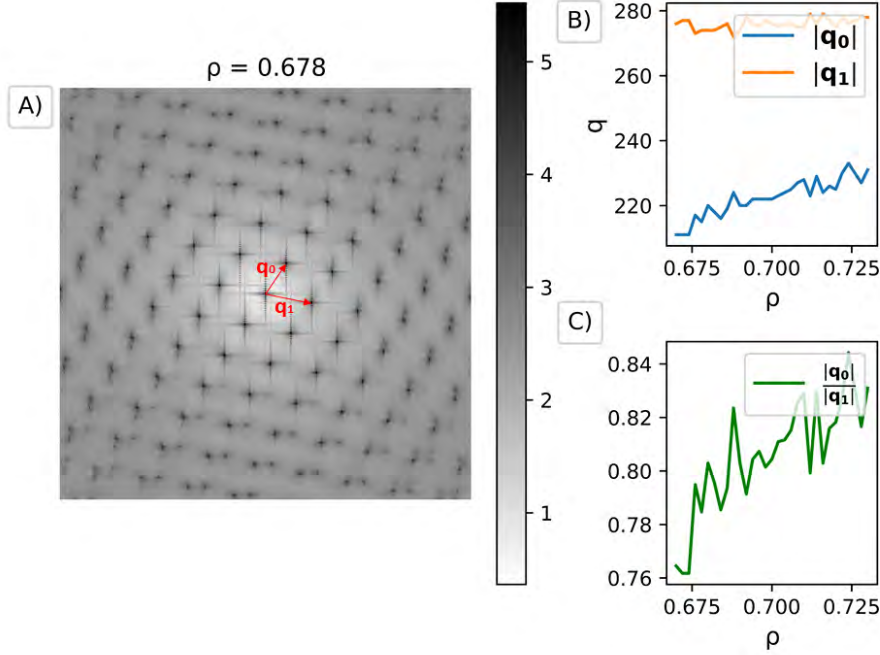

**Figure 7**. (A) Structure function of the center of masses of pairs of particles in the pairs phase, ρ = 0.678, T = 0.005, in the Truskett system of 1200 particles. $\mathbf{q_0}$ and $\mathbf{q_1}$ represent the two vectors that comprise the stretched hexagonal shape of the first diffraction ring. (B) Plot of $\mathbf{q_0}$ and $\mathbf{q_1}$ magnitudes ($|\mathbf{q_0}|, |\mathbf{q_1}|$) as a function of density along the T = 0.005 isotherm in the Truskett system. (C) Plot of the ratio of $|\mathbf{q_0}|$ to $|\mathbf{q_1}|$ as a function of density along the T = 0.005 isotherm in the Truskett system. We note that the stretching of the hexagonal diffraction ring, $|\mathbf{q_0}|/|\mathbf{q_1}|$, increases sub-linearly as density is increased.

Increasing the density further generates a smooth transition from the pairs phase to the string phase, and when the density is increased from ρ = 1.100 to ρ = 1.236, the system is gradually transformed from the string phase to the Kagome phase via particles within the string phase encroaching into the regions between the chains. Lastly, there is a loop in the pressure-density isotherm at around ρ = 1.580, representing the first order transition from the Kagome phase to the high-density hexagonal solid phase with clear phase-coexistence.

Pair correlation functions for the several phases identified are displayed in Fig. 8. A more detailed representation of the pair correlation function in the string phase, and its decay, is displayed in Fig. 9. We note that the envelope of the pair correlation function for the string phase decays algebraically, the signature of quasi-long-range translational order.

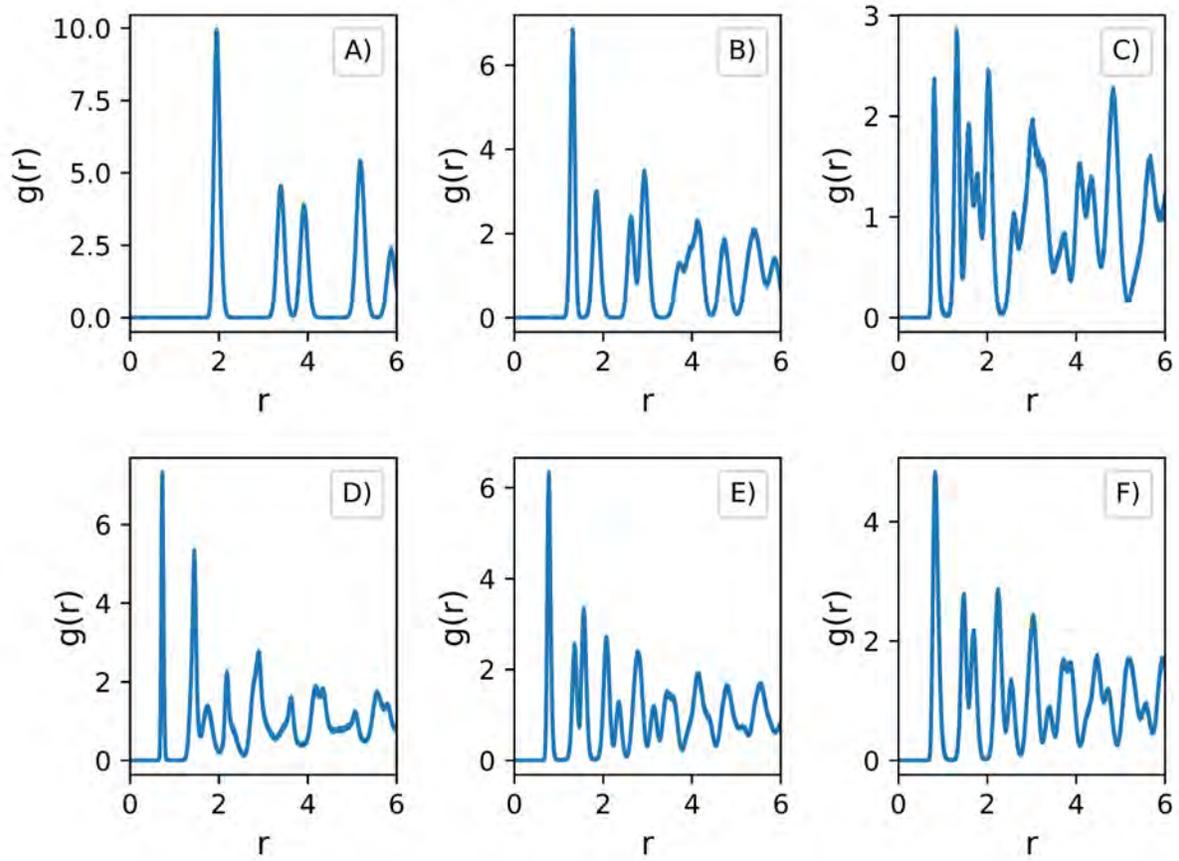

**Figure 8.** Pair correlation functions for the Truskett system. (A) the low-density triangular solid phase (T = 0.005, ρ = 0.300, N = 1200); (B) the square solid phase (T = 0.005, ρ = 0.580, N = 1200); (C) the pairs phase (T = 0.005, ρ = 0.670, N = 1200); (D) the string phase (T = 0.005, ρ = 1.000, N = 1200); (E) the Kagome solid phase (T = 0.005, ρ = 1.400, N = 1200); and (F) the high-density triangular solid phase (T = 0.005, ρ = 1.598, N = 1200).

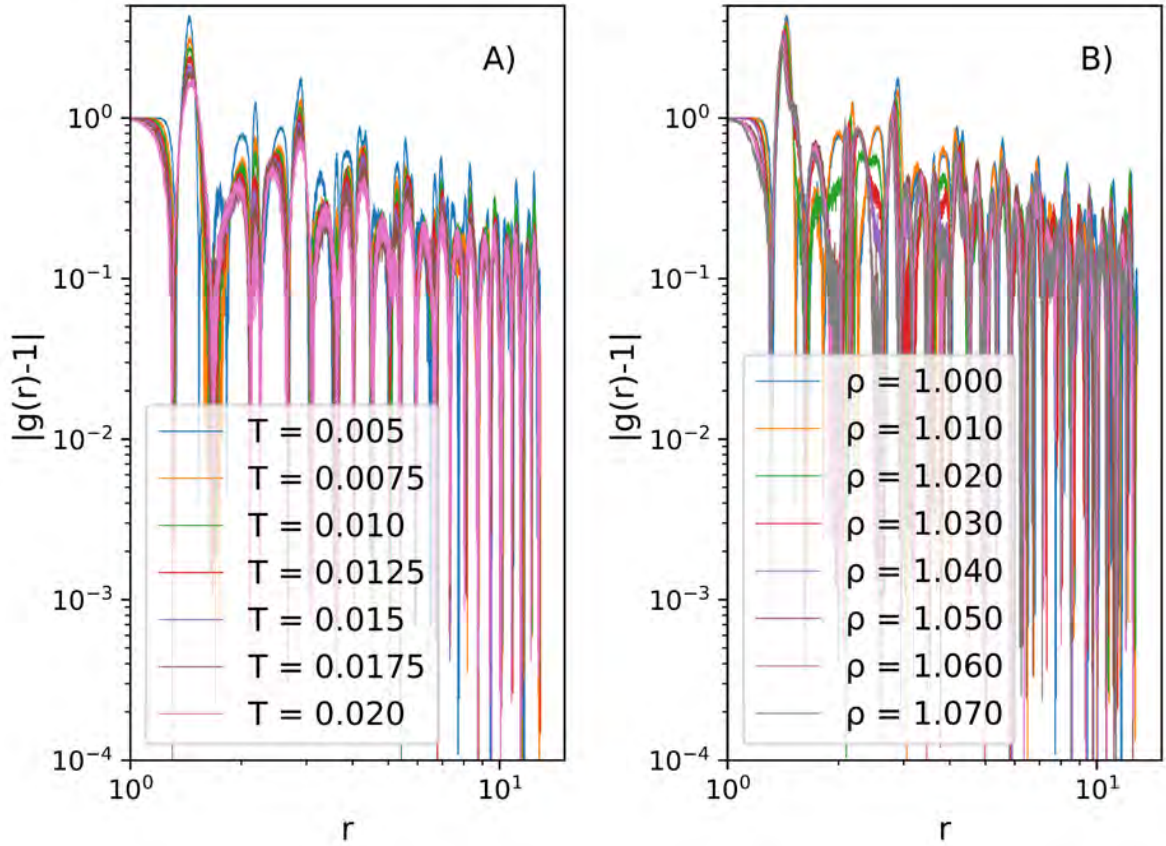

**Figure 9.** A) Pair correlation function in the string phase with ρ = 1.000 at several temperatures, N =1200, and B) pair correlation function in the string phase at several densities around at T = 0.005, N =1200.

We are not aware of any prior report of a phase consisting of spontaneously formed dimers in any of the 2D systems that exhibit a string phase. There is a report [6], that we will discuss later, of an ordered 2D phase of permanent dimers in a system that also has a string phase. Given that the Truskett pair interaction is monotonic repulsive, why do pairs of particles become stable and why is that stability a function of density? We find, from the pair correlation function for the Truskett system displayed in Fig. 6C and the pair force function displayed in Fig.1 that the center-to-center intra particle distance in a pair is close to the threshold where the pair force function starts to rapidly increase (r/σ ≈ 0.7), and the distance between adjacent particles in different pairs is close to the inflection point where the force starts to drop from its plateau value to zero (r/σ ≈ 1.2). Clearly, there is a range of intra-pair separation, $0.7 < \frac{r}{\sigma} <$ 1.2, with nearly constant force between the particles. The interaction between particles in different pairs, when $\frac{r}{\sigma} > 1.2$, generates a force that is smaller than the intra-particle force. And, to the extent that pairs are arranged almost linearly as seen in Fig. 4F, adjacent distinct pairs experience an increase in repulsive force when they approach $\frac{r}{\sigma} \approx 1.2$. We suggest that pair formation is entropically driven by the tradeoff of an increase in range of motion of the centers of mass of pairs for a decrease in intra-particle motion in a pair. The tradeoff will fail when, as a consequence of increased density, the motion of the centers of mass of pairs is sufficiently restricted by decrease in the center of mass separation. Then it is favorable for the separated pairs to merge and become a uniformly spaced string of particles.

The string phase in the Truskett system emerges from the pairs phase when the density is increased. The formation of a string phase in a 2D system has been reported for several different monotonic repulsive central pair potentials. For example, the 2D system with a hard disc plus rectangular step repulsion (so-called core-corona potential), studied by Pattabhiraman and Dijkstra [3], exhibits a string phase with overlapping coronas of nearest neighbor particles on the same string and no overlap of coronas of particles on different neighbor strings. In this case, with discontinuities in the pair potential at the core and at the edge of the corona, there is a convincing argument that string formation is accounted for by minimization of the system potential energy. The overlap of the corona of one particle with that of two other particles in the same string is energetically favorable relative to that associated with the overlap of the corona of one particle with more than two other coronas, as happens if the strings are not separated. The smooth monotonic decline of the Truskett potential with increasing pair separation, displayed in Fig. 1, is very different from the shape of the core-corona potential, and the argument vis a vis string stabilization by minimization of the potential energy is harder to make. Nevertheless, the arrangement of particles in the Truskett system string phase appears to be sensibly the same as that in the core-corona system strings phase [3]. A difference is that the strings in the Truskett system span the simulation cell at all densities whereas those in the core-corona system start as small chains that, with increasing density, grow to span the simulation cell. The pair correlation function for the string phase, with $\rho = 1.000$ at $T = 0.005$, when compared with the Truskett pair force function shown in Fig. 1, reveals that the nearest neighbor peak corresponds to a particle separation along the string that is very close to the separation at which the pair force starts to become strongly repulsive (there is no minimum in the pair force function) and that the next peak corresponds to separation of the strings at a distance that is the end of the plateau in the pair force. The next two peaks in the pair correlation function correspond to the separation of a particle from second and third neighbors along the same string.

## 2. 2D Phases Supported by the Torquato Pair Interaction

Simulations similar to those reported in the previous Section were carried out for 2D systems with the Torquato pair interaction. A set of superimposed pressure-density isotherms is displayed in Fig. 10. We find that the Torquato system passes through a sequence of phases similar to those in the Truskett system, namely:

Liquid → Low-Density Hexagonal → String → Kagome → High-Density Hexagonal.

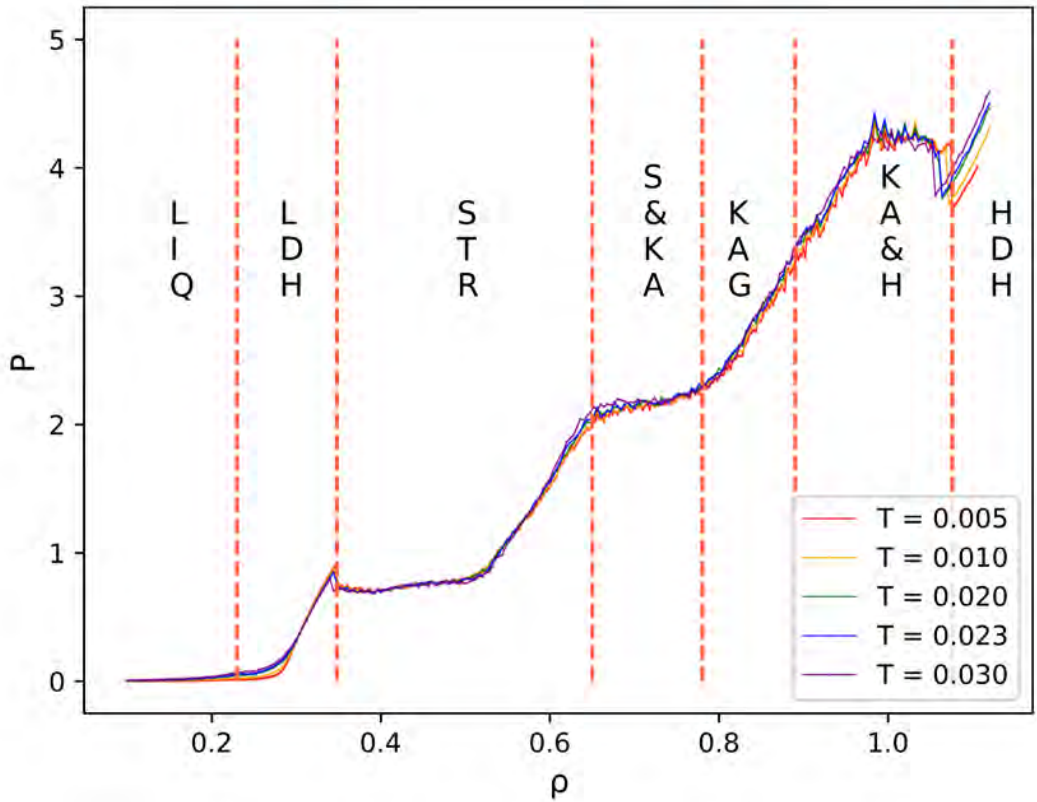

**Figure 10.** Superimposed pressure-density isotherms for the Torquato system with 1200 particles. LIQ = liquid, LDH = low-density hexagonal solid, STR = string, S&KA = coexistence

of string phase and Kagome lattice, KAG = Kagome, KA&H = coexistence of Kagome phase and high-density hexagonal solid, HDH = high-density hexagonal solid.

The particle configurations in Figs. 11 and 12 suggest that along the T = 0.005 isotherm, the Torquato system begins in a liquid phase at low density before transitioning into a low-density hexagonal solid phase. This hexagonal solid phase undergoes a first-order transition into a phase with paired particles, and these pairs then grow into longer chains as the density is increased. The chains then transform into the Kagome phase via a process that is slightly different from that found in the Truskett system. The winding nature of the chains in the Torquato system allows for pockets to form via select particles entering into the region between the chains. This differs from the transition into the Kagome phase in the Truskett system, during which the particles seem to collectively and gradually encroach into the region between the chains. The sequence of transitions described above is reflected in the superimposed pressure-density plots in Fig. 10. There, one observes loops at $\rho = 0.348$ and $\rho = 1.076$ where the hexagonal solid to pairs phase and the Kagome phase to high-density hexgonal solid phase transitions occur, respectively. These are first-order transitions, as suggested by the phase coexistence seen in the images.

The string phase in the Torquato system first appears with short strings that then grow to span the simulation cell. It transforms, at higher density, to a hexagonal phase. This behavior is like that in the core-corona system. Indeed, the shape of the Torquato pair interaction resembles a softened core-corona interaction in that there is a strong increase in repulsion at a rather well-defined particle separation, followed by a near plateau in the interaction as separation is increased, and then a decay of the interaction to zero.

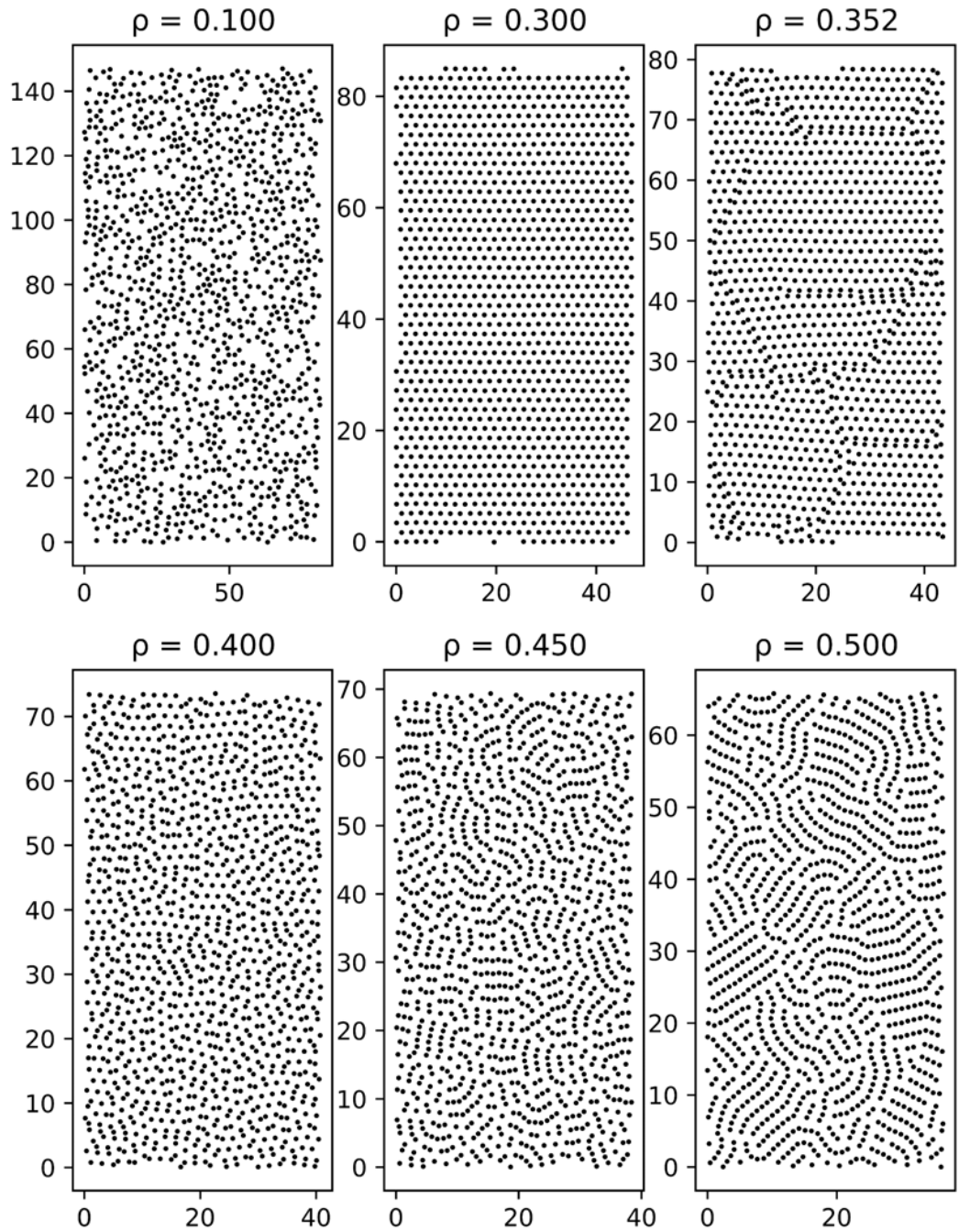

**Figure 11.** Snapshots of particles in the Torquato system with 1200 particles at different densities and T = 0.005.

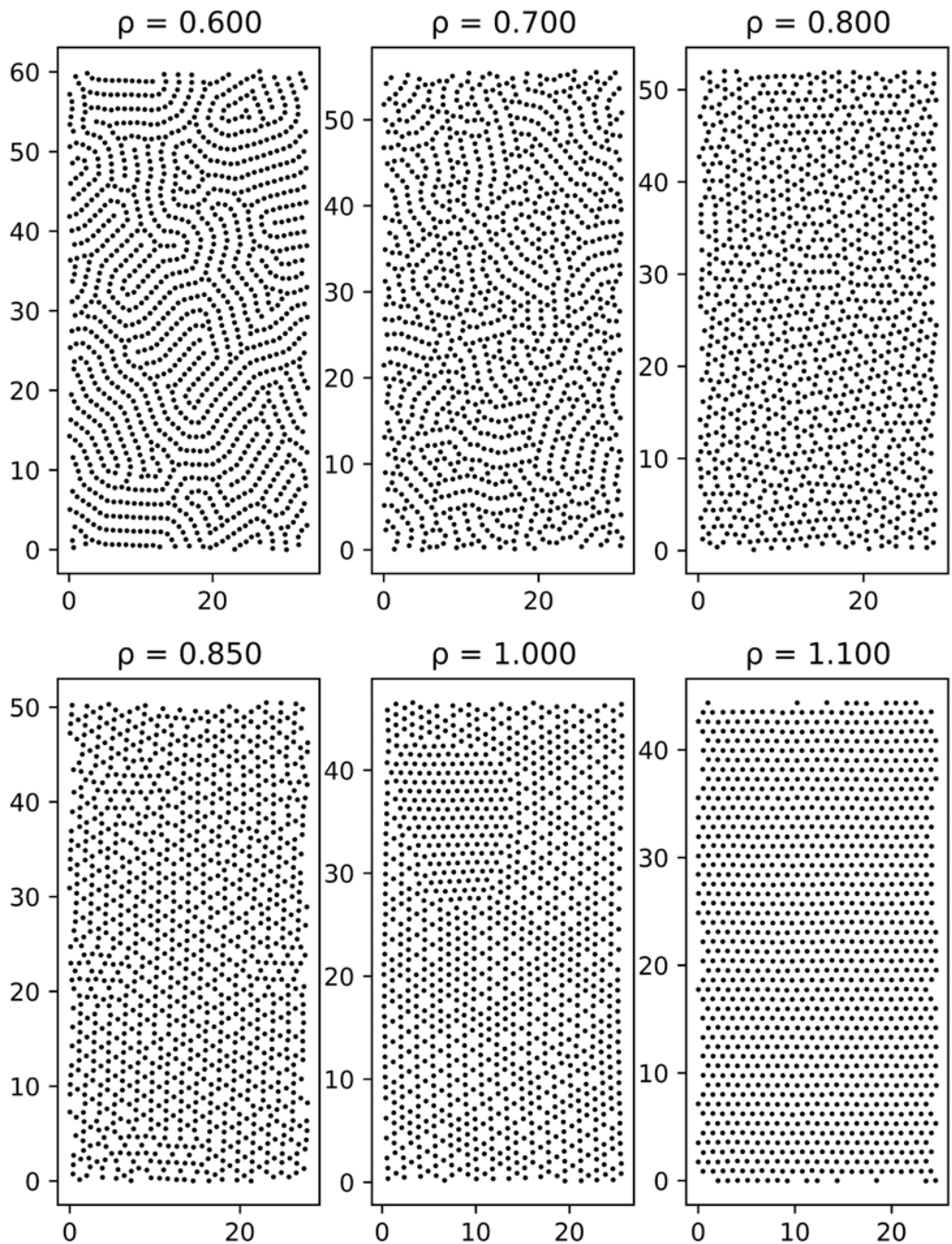

**Figure 12.** Snapshots of particles in the Torquato system with 1200 particles at different densities and T = 0.005.

## 3. Structured Fluctuations Near the Liquid-Kagome and the String-Kagome Phase Boundaries

As noted in the Introduction, previous studies of structured fluctuations in the liquid phases of 2D systems that support a variety of ordered solids have revealed that they exhibit the same symmetries as the respective ordered phases in those systems; no other symmetries have been observed to occur. Part of the motivation for the research reported in this paper is our interest in learning what precursor transient fluctuations accompany the formation of an "open" structure such as a Kagome or honeycomb lattice. None of the systems previously studied has an ordered solid phase with Kagome or honeycomb lattice structure. Motivated in part by the expectation of parallel behavior vis a vis structured fluctuations in the other systems studied, and in part by the intuitive notion that structured fluctuations with an open packing motif must compete with unstructured fluctuations with the same local density, we now use the results of the simulations reported in this paper to ask what transient ordered fluctuations in the liquid are associated with a transition from the liquid state to a solid with a Kagome lattice? Further noting that both the Truskett and Torquato potentials support a string-to-Kagome phase transition we also ask what transient ordered fluctuations in the string phase are associated with that transition.

We consider first the string-to-liquid and liquid-to-Kagome transition in a system with the Truskett interaction. To do so we extended our simulations of the Truskett system to higher temperature than discussed above. Along this higher temperature isotherm (T = 0.023), the Truskett system passes through the following sequence of phases:

Liquid #1 → Low-Density Hexagonal → String → Liquid #2 → Kagome → Liquid #3.

The absence of a pairs phase and square phase when T = 0.023 is similar to the sequence of phases along the T =0.005 isotherm for the Torquato system, with the exception of the second and third liquid phases in the Truskett system. Fig. 13 indicates the densities at which this new Liquid #2 phase is entered from the string phase and exited into the Kagome phase. Here, the Liquid phase #2 in the T= 0.023 Truskett system succeeds the string phase at $\rho$ = 1.052, as visualized in the diffraction patterns of Fig. 14. A Liquid #2-to-Kagome transition then occurs around $\rho$ = 1.252 (see Fig. 15). Along this isotherm, the Truskett system subsequently undergoes a second transition from the Kagome phase into an even higher density liquid phase (Liquid #3) at ρ = 1.414.

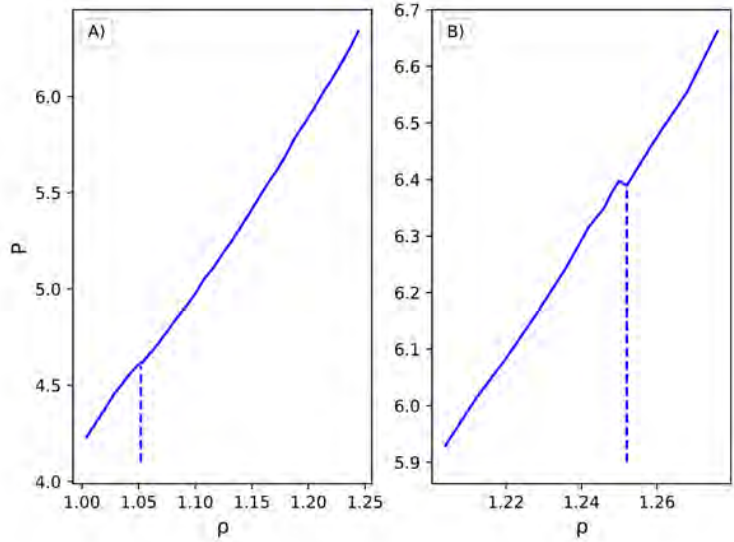

**Figure 13.** Pressure-density isotherms at T = 0.023 for the Truskett system with 1200 particles. The dashed line in A marks the phase transition from the string phase to the Liquid #2 phase. The dashed line in B marks the phase transition from the Liquid #2 phase to the Kagome phase.

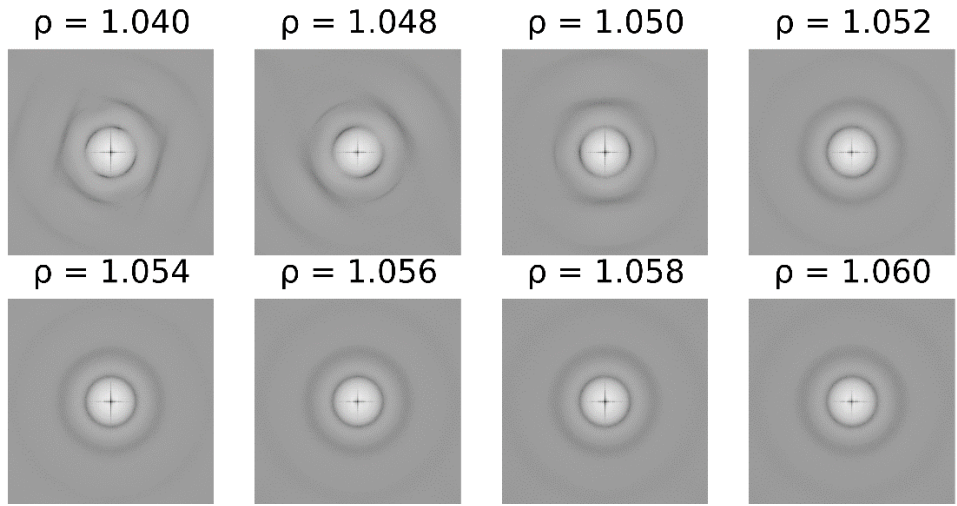

**Figure 14.** Diffraction patterns calculated along the T = 0.023 isotherm for the 11250-particle Truskett system. For ρ ≥ 1.052 the system is in the high-density liquid phase. At lower densities the system is in the string phase.

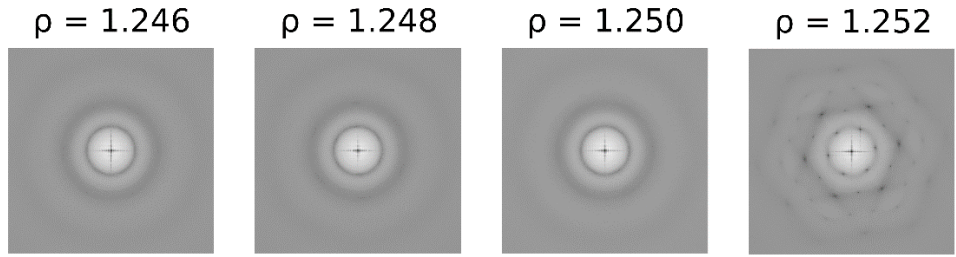

**Figure 15.** Diffraction patterns calculated along the T = 0.023 isotherm for the 11250-particle Truskett system. For $\rho \leq 1.250$ the system is in the high-density liquid phase. At higher densities the system is in the Kagome phase.

Evidence for the emergence of the Kagome-to-Liquid #3 transition is also found in Figs. 16 and 17. Fig. 16 displays a set of such higher temperature pressure-density isotherms for the Truskett system. We note that as the temperature increases the Kagome-to-high-density hexagonal solid first-order transition is shifted to lower densities until it disappears along the isotherms T = 0.023 and T = 0.030, as verified by the diffraction patterns shown in Fig. 16. When T = 0.23 this transition occurs at about $\rho$ = 1.414. A similar search of the Torquato system phase diagram along the isotherms T = 0.005, 0.010, 0.023, and 0.030 failed to find a Kagome-to-liquid phase transition.

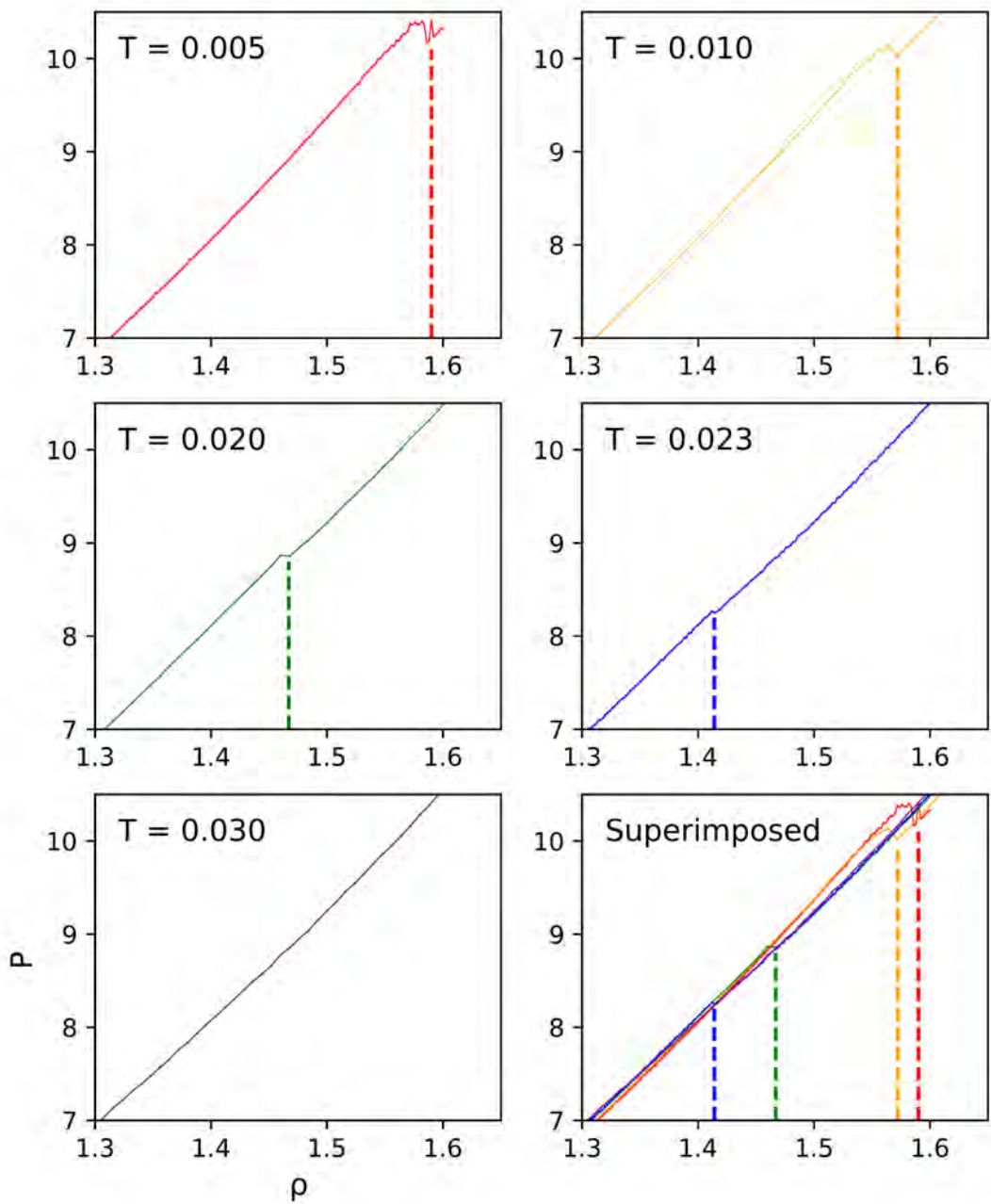

**Figure 16.** Pressure-density isotherms for the Truskett system with 1200 particles with T = 0.005, 0.010, 0.023, and 0.030.

Indeed, for $\rho = 1.450$ the intensity distribution approaches that characteristic of hexagonal lattice symmetry. The simplest interpretation of this observation is that as the density of the liquid isothermally approaches the transition density weak transient structured fluctuations with Kagome symmetry appear, in coexistence with the transient hexagonal symmetry fluctuations characteristic of the liquid, eventually replacing those with hexagonal symmetry. This transformation of the symmetry of the transient ordered fluctuations occurs in a density range of order 1- 2% from the transition density.

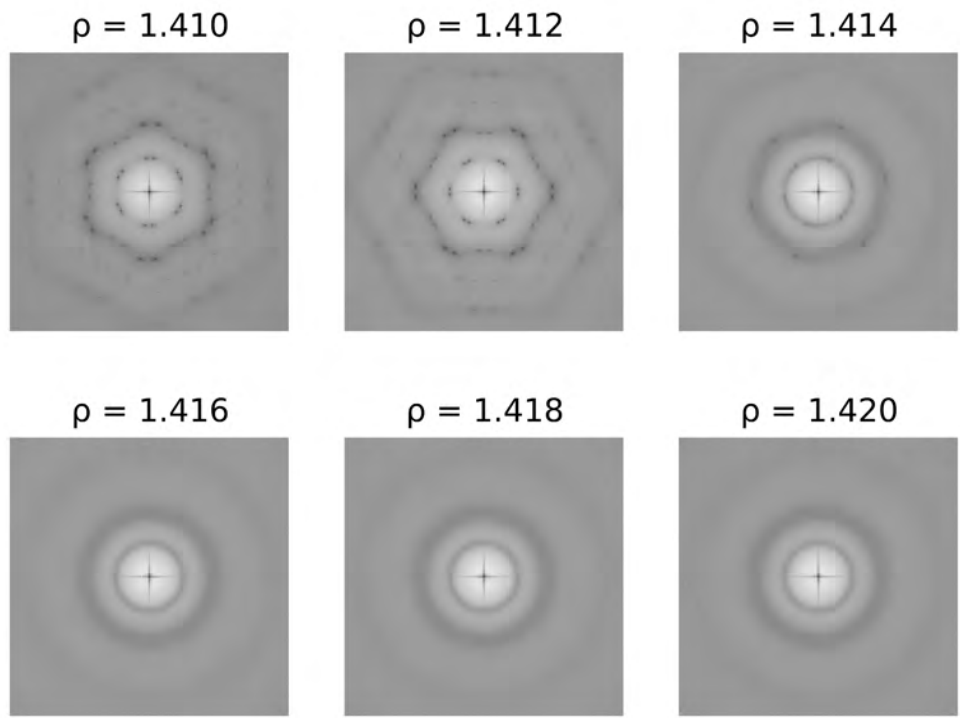

**Figure 17.** Diffraction patterns calculated along the T = 0.023 isotherm for the 11250-particle Truskett system. For $\rho \geq 1.414$, the system is in the high-density liquid phase. At lower densities the system is in the Kagome phase.

The presence of two liquid phases adjacent to the Kagome phase in the Truskett system at T = 0.023 offers an opportunity to compare fluctuations in a liquid phase as they might relate to both an open packed system and the string phase. Using the ACCF, we first examined fluctuations in the Liquid #2 phase near the string-to-liquid transition and liquid-to-Kagome transition. The ACCF's in Fig. 18 were calculated along the first diffraction ring of the Kagome lattice. Fig. 18B, indicates the growth of six-fold peaks as the Kagome phase is approached. These peaks at 60º and 120º are faintly mimicked in the liquid phase near the string-to-liquid boundary, as shown in Fig. 18A. As mentioned earlier, six-fold peaks in an ACCF calculated along the first diffraction ring may be associated with fluctuations of hexagonal packing or Kagome packing. Therefore, in order to identify a clearer signature of Kagome fluctuations one must compare a cross correlation of the intensities at two scattering vectors along the second diffraction ring. Such calculations are explored later in this section. Nevertheless, the six-fold peaks in Fig. 18 still suggest that there exist structured fluctuations in the Liquid #2 phase of the T = 0.023 Truskett system.

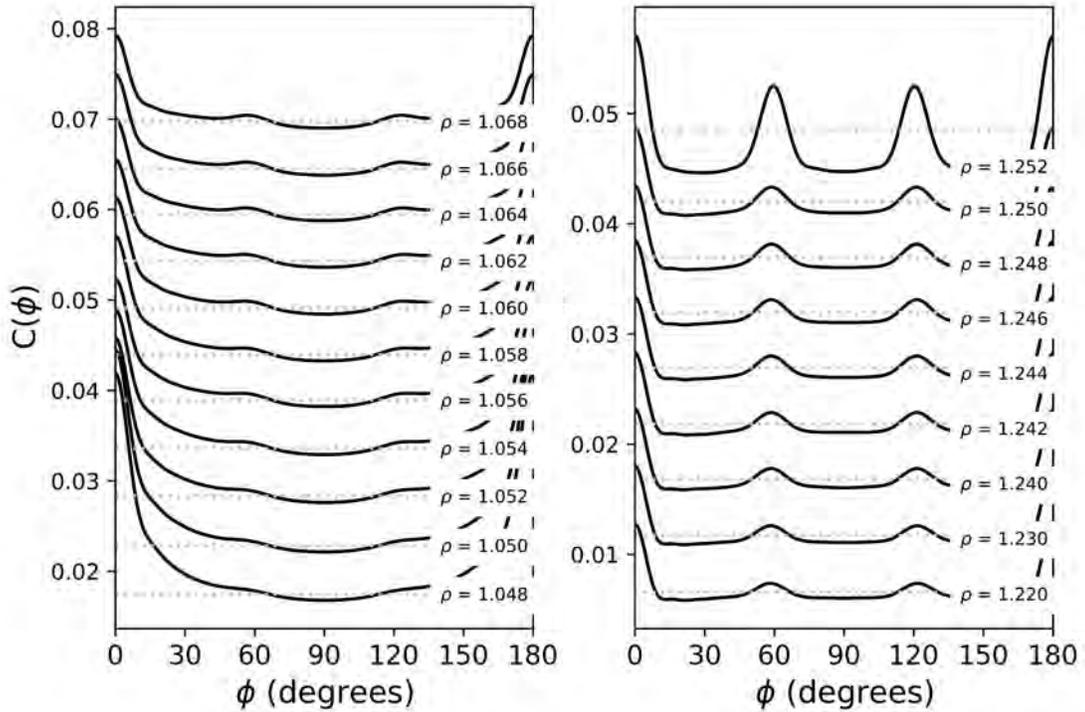

**Figure 18.** The ACCF line-shapes cross-correlating intensities at the scattering vector of the first diffraction ring ($|\mathbf{q_0}|$, arbitrary units). These are produced within the string, Liquid #2, and Kagome phases of the T = 0.023 isotherm at densities $1.048 \leq \rho \leq 1.252$ for the 11250-particle Truskett system.

Increasing the temperature from T = 0.005 to T = 0.023 induces the emergence of the Liquid #2 phase ($1.052 \leq \rho \leq 1.250$) in the Truskett system. Given that this liquid phase spans the density range of the string phase and pockets phase at T = 0.005, and given that the pockets phase may be considered an area of string-Kagome coexistence, we expected that evidence of six-fold order might be found along the string-to-Kagome transition not only in the Truskett system but also the Torquato system. As mentioned earlier, Figs. 5 and 12 already convey that the string phase of the Truskett system is different in form from that of the Torquato system. At T = 0.005, the particle array in the Truskett string phase is more linear, and in the Torquato string phase is more coiled. Moreover, these alternative string phases approach the Kagome phase in dissimilar fashions. Examination of the particle configurations generated by compression of the Truskett system at T = 0.005 suggests that the Truskett string-to-Kagome transformation pathway includes intermediate stages with local particle arrangements with honeycomb symmetry (Fig. 19C). As the area of the simulation box is decreased, alternating particles along the string move inward in a zig-zag pattern that then becomes honeycomb packing. Upon further compression, this honeycomb packing buckles into the Kagome lattice. In contrast, examination of the particle configurations generated by compression of the Torquato system suggests that the Kagome phase is formed by coiled strings merging together at three-particle joining sites (Fig. 19H). The increased concentration of these joining sites as the density increases generates a Kagome phase with imperfections such as 8-particle rings (Fig. 19J).

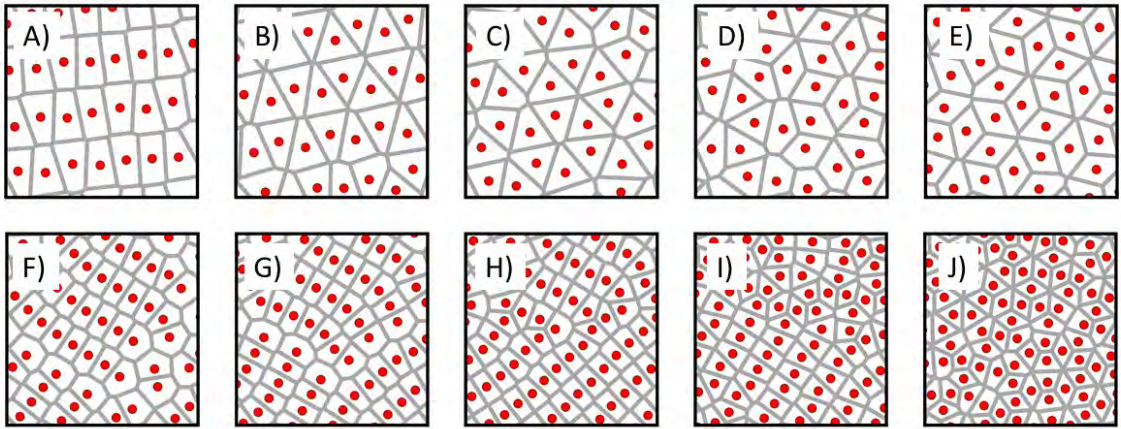

**Figure 19.** Particle configurations generated by compression of the Truskett system (A-E) and Torquato system (F-J) along the path from the string phase to the Kagome phase. The red dots mark the locations of the particle centers and are not representative of particle size. The gray lines show the Voronoi tessellations of the particle configurations.

Using the ACCF, we offer a more quantitative picture of the pockets in the Truskett system and the inter-string particles in the Torquato system that give way to the Kagome phase. Fig. 20 suggests that the Truskett system's pockets phase has stronger six-fold transient order than the inter-string particles in the Torquato system. The presence of ACCF peaks at angular separations other than 60º and 120 º in both systems' string-to-Kagome density range is unambiguous evidence that in both systems there exist transient local configurations with other than six-fold symmetry, noting that the narrower peaks in the Truskett ACCF's indicate greater preference in the that system for six-fold symmetry. The Torquato system's ACCF's, such as when ρ = 0.720, suggest that the transition path into the Kagome phase—via coiled strings with increasingly more inter-chain particles—involves many non-six-fold particle configurations. Together, the ACCF calculations and Voronoi tessellations for the Truskett and Torquato systems suggest that the pathways utilized by the two systems in their respective transition to the Kagome phase are different. One possible determinant of the different mechanisms is the uncoiled vs. coiled nature of the preceding string phases. However, a recent paper by Zhu, Truskett, and Bonnecaze offers one figure that shows a string phase similar to the Torquato string phase transitioning into a honeycomb phase [7]. At present we cannot determine whether or not the transition from a string phase to a Kagome phase will always involve intermediate particle configurations with honeycomb symmetry.

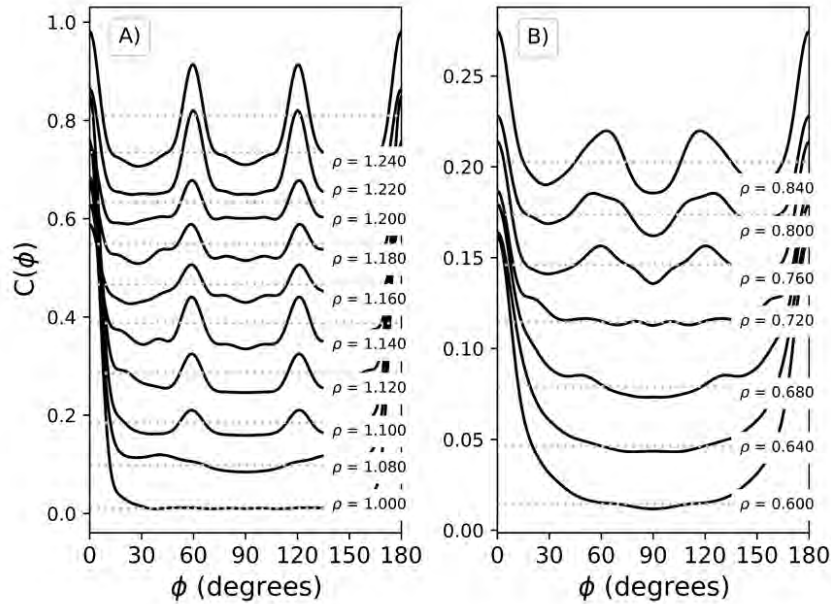

**Figure 20.** The ACCF line-shapes cross-correlating intensities at the scattering vector of the first diffraction ring ($|\mathbf{q_0}|$, arbitrary units). Plot A displays ACCF's from the 1200-particle Truskett

system at T = 0.005 in the string, pockets, and Kagome phases. Plot B displays ACCF's from the 1200-particle Torquato system at T = 0.005 in the string, coexistence, and Kagome phases.

We have also calculated the ACCF's for the Truskett system with T = 0.023 for the density range $1.400 \leq \rho \leq 1.450$, bracketing the Kagome lattice-to-liquid transition at $\rho = 1.414$. Fig. 21 displays these ACCF's for values of **k** and **q** corresponding to the circles that pass through the vertices and the midpoints between the vertices of the second ring of a Kagome lattice diffraction pattern. The ACCF's for densities $\rho \leq 1.414$, i.e. in the ordered solid, show clearly the alteration of intensity of diffraction spots in the second ring that is characteristic of Kagome packing. The ACCF's for densities $\rho > 1.414$, i.e. in the liquid, show a clear but much weaker persistence of this pattern, thereby identifying transient ordered fluctuations in the liquid very close to the liquid-Kagome transition with the same structural motif as the solid. We note that the intensity of the diffraction spots at the vertices relative to that at the diffraction spots midway between the vertices decreases as the density moves away from the transition density.

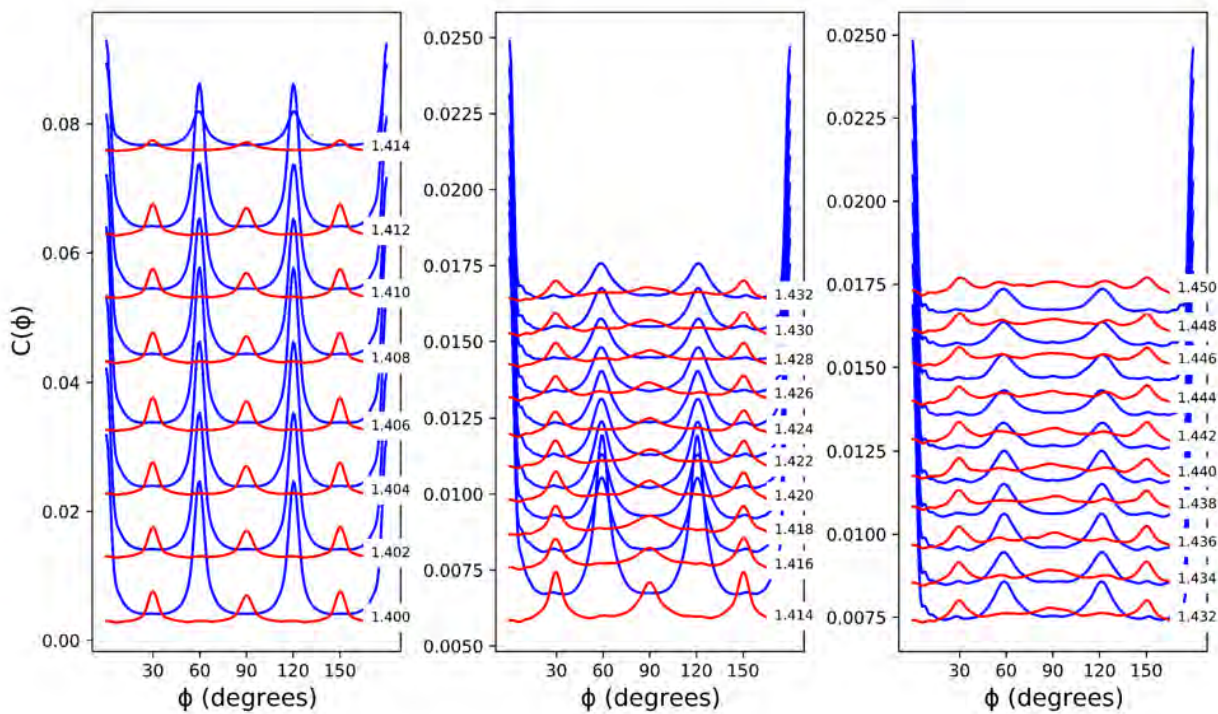

**Figure 21.** The ACCF line-shapes cross-correlating intensities at two different scattering vectors corresponding to the circles that pass, respectively, through the vertices and through the midpoints between the vertices of the second ring of a Kagome diffraction pattern. These are generated from the simulations along the isotherm T = 0.023 at densities $1.400 \leq \rho \leq 1.450$ for the 11250-particle Truskett system. The blue line-shapes indicate cross-correlation with $|\mathbf{q}| = 2|\mathbf{q_0}|$ and $\mathbf{k} = 2|\mathbf{q_0}|$. The red line-shapes indicate the cross-correlation with $|\mathbf{q}| = \frac{5}{3}|\mathbf{q_0}|$ and $|\mathbf{k}| = 2|\mathbf{q_0}|$. Here, $|\mathbf{q_0}|$ is the magnitude of the first diffraction ring of the Kagome lattice with arbitrary units. Note the different vertical scale used in the left panel and used in both the central and right panels.

## Discussion

In this paper we have addressed two questions:

Do different 2D systems, each composed of particles with monotonic isotropic everywhere repulsive pair potential, exhibit the same sequence of phase transitions as the density is isothermally increased?

Do transient structured fluctuations with Kagome structure develop prior to transition from a liquid to a Kagome lattice? From an ordered string phase to a Kagome lattice?

With respect to the first question, it is revealing to compare our findings for 2D systems with the Truskett (Eq. (1)) and Torquato (Eq. (2)) potentials with those of other investigations of 2D systems with other pair potentials, specifically the Hertzian potential [5]

$$u_H = \epsilon\left(1 - \frac{r}{\sigma}\right)^{5/2}, \qquad (3)$$

the Lennard-Jones-Yukawa (LJY) potential [6]

$$u_{LJY} = 4\epsilon\left[\left(\frac{\sigma}{r}\right)^{12} - \left(\frac{\sigma}{r}\right)^6\right] + \frac{A}{r}e^{-\kappa r} \quad \text{for } \frac{r}{\sigma} \leq 6;\ u_{LJY} = 0 \text{ for } \frac{r}{\sigma} \geq 6, \qquad (4)$$

the Core-Corona potential [3]

$$u_{CC} = \begin{cases} \infty, & r \leq \sigma \\ \varepsilon, & \sigma < r \leq \delta, \\ 0, & r > \delta \end{cases} \qquad (5)$$

for several values of $\frac{\delta}{\sigma}$, and the Daoud–Cotton model potential [7]

$$\frac{u_{DC}}{k_B T} = \frac{5}{8} f^{\frac{3}{2}} \begin{cases} -\ln\left(\frac{r}{\sigma}\right) + \frac{1}{1+\sqrt{\frac{f}{2}}}, & r \leq \sigma \\ \frac{\sigma \exp\left[-\frac{\sqrt{f}(r-\sigma)}{2\sigma}\right]}{r\left(1+\sqrt{\frac{f}{2}}\right)}, & r > \sigma \end{cases} \quad (6)$$

for several values of $f$. For this set of pair interactions, along a suitably chosen low-temperature isotherm that in each system passes through all of the phases, one observes the following sequences of phases:

**Truskett** [1]: Liquid → Hexagonal → Square → Pairs → String → Kagome → Hexagonal

**Torquato** [2]: Liquid → Hexagonal → String → Kagome → Hexagonal

**Hertzian** [5]: Liquid → Hexagonal → Square → Pentagonal → Stretched Hexagonal → Hexagonal → Square → Rhombohedral → Stretched Hexagonal → Hexagonal

**LJY** [6]: Liquid → Hexagonal → String → Hexagonal → Kagome → Hexagonal

**Core-Corona** ($\frac{\delta}{\sigma}$ =1.95) [3]: Liquid → Hexagonal → Liquid → String → Honeycomb → Hexagonal → Sigma → Hexagonal

**Core-Corona** ($\frac{\delta}{\sigma}$ =1.40) [3]: Liquid → Hexagonal → Liquid → Square → Quasi-crystal → Hexagonal

**Core-Corona** ($\frac{\delta}{\sigma}$ =1.27) [3]: Liquid → Hexagonal → Liquid → Rhombus → Quasi-crystal → Hexagonal

**Daoud–Cotton** ($f$ = 50, 68, 150) [7]: Hexagonal → Stripe → Honeycomb → Kagome → Hexagonal

All of these pair potentials are everywhere repulsive. For all of these potentials, along a low-temperature isotherm that passes through all of the phases of the system the structure of the lowest density ordered phase is hexagonal, as is the structure of the highest density phase. The sequence of higher density phase transitions, String → Kagome → Hexagonal, is common to the systems with Truskett, Torquato and LJY interactions, while the system with Hertzian potential exhibits a richer sequence of phases and phase transitions. The sequences of phases for systems with the Truskett and Torquato interactions differs from that of the system with the LJY interaction via the appearance, in the former cases, of phases with Square, Pair and Strings structures. All of the pair interactions cited above refer to unstructured monomeric particles [1-7]. Surprisingly, simulation studies by Yang et al reveal that the phase diagram of a 2D system composed of dumbbell particles (two particles connected by a rigid bond of length σ), consists of a similar sequence of structures, including a stripe phase [6]. Their dumbbells are comprised of two particles connected by a rigid bond of length σ, with all particles interacting via the isotropic LJY potential function displayed in Eq. (4). The sequence of transitions found for the dumbbell system from low density to high density along the T = 0.005 isotherm is [6]:

**LJY dimer**: Triangular → Stripe-Triangular → Hexagonal → Kagome → coexistence → Triangular

An interesting feature of the "stripe-triangular" phase is that, towards the high-density end of the range of stability of this phase, the stripe pattern can be superimposed on a triangular lattice when the particle separation is less than that corresponding to the "minimum repulsive force" well (1.4 σ) between the particles [6].

The commonalities in the phase diagrams of the several 2D systems described suggests the existence of a universal yet still not fully understood mechanism driving all to favor a similar series of packing arrangements as the density is increased. The characteristics of a pair potential that supports, as a function of increasing density, many ground state lattices, has been a subject of interest for some time. An important theorem by Sütő [18, 19] states that if the Fourier Transform of the pair potential is non-negative and decays to zero at a wave number greater than $K_0$, then at a dimension-dependent density $\rho_d$ approximately below $K_0^d$, the system supports an infinite number of unique periodic ground state configurations. At densities above $K_0^d$, the ground state configurations are continuously degenerate [18, 19]. Considering the several pair potentials we have discussed, we note that the Truskett, Torquato, LJY, and Core-Corona potentials have the following qualitative differences from the Hertzian potential: The former four monotonic repulsive interactions have a shoulder at some finite r whereas the Hertzian potential does not, and the former four interactions diverge at the origin, whereas the Hertzian potential is finite at the origin. Meanwhile, the Daoud–Cotton potential does not have a shoulder like the Truskett, Torquato, LJY, and Core-Corona potentials; however, it does diverge at the origin. Consequently, the Truskett, Torquato, LJY, Core-Corona and Daoud–Cotton potentials do not have a Fourier Transform, whereas the Hertzian potential does. We infer that satisfying the Süto theorem is not a necessary condition for the support of multiple distinct lattice structures by a particular pair potential.

## Acknowledgements

The research reported in this paper was primarily supported by the University of Chicago Materials Research Science and Engineering Center, funded by National Science Foundation Grant No. DMR-1420709.

# References

[1] W. D. Piñeros, M. Baldea, and T. M. Truskett, J. Chem. Phys. **145**, 054901 (2016).

[2] G. Zhang, F. H. Stillinger, and S. Torquato, Physical Review E **88**, 042309 (2013).

[3] H. Pattabhiraman and M. Dijkstra, Soft Matter **13**, 4418 (2017).

[4] T. Dotera, T. Oshiro, and P. Ziherl, Nature **506**, 208 (2014).

[5] Yu. D. Fomin, E. A. Gaiduk, E. N. Tiosk, and V. N. Ryzhov, Mol. Phys. **16**, 3258 (2018).

[6] Z. Yang, M. Dutt, and Y. C. Chiew, Materials Research Express **6**, 075076 (2019).

[7] Xilan Zhu, T. M. Truskett, and R. T. Bonnecaze, Soft Matter **15**, 4162 (2019).

[8] B. J. Ackerson, T. W. Taylor, and N. A. Clark, Physical Review A **31**, 3183 (1985).

[9] J. Steinhardt, D. R. Nelson, and M. Ronchetti, Phys. Rev. Lett. **47**, 1297 (1981).

[10] L. V. Woodcock, Z. Naturforsch. **26a**, 287 (1970).

[11] A. Sheu and S. A. Rice, J. Chem. Phys. **128**, 244517 (2008).

[12] Z. Krebs, A. B. Roitman, L. M. Nowck, C. Liepold, B. Lin and S. A. Rice, J. Chem. Phys., **149**, 034503 (2018).

[13] D. E. Dudalov, E. N. Tsiok, Y. D. Fomin, and V. N. Ryzhov, J. Chem. Phys. **141**, 18C522 (2014).

[14] D. E. Dudalov, Y. D. Fomin, E. N. Tiosk and V. N. Ryzhov, Soft Matter **10**, 4966 (2014).

[15] N. P. Kryuchkov, S. O. Yurchenko, Y. D. Fomin, E. N. Tsiok, and V. N. Ryzhov, arXiv:1712.04707 [cond-mat.soft] (2017).

[16] S. Plimpton, J Comp Phys **117**, 1 (1995), http://lammps.sandia.gov.

[17] W. Humphrey, A. Dalke, and K. Schulten, J. Molec. Graphics **14**, 33 (1996), http://www.ks.uiuc.edu/Research/vmd/.

[18] A. Süto, Physical Letters Review **95**, 265501 (2005).

[19] A. Süto, Commun. Math. Phys. **305**, 657 (2011).